\documentclass[useAMS,usenatbib]{mnras}

\usepackage{graphicx}

\title[Polarization of broad line emission from AGNs]{Polarization of broad line emission from  AGNs with determined virial factors}

\author[M.Yu. Piotrovich, Yu.N. Gnedin,  T.M. Natsvlishvili, S.D. Buliga]{M.Yu. Piotrovich \thanks{E-mail: mpiotrovich@mail.ru}, Yu.N. Gnedin \thanks{E-mail: gnedin@gao.spb.ru}, T.M. Natsvlishvili, S.D. Buliga \\
Central Astronomical Observatory at Pulkovo, 196140, Saint-Petersburg, Russia}

\date{Accepted for publication in MNRAS.}

\pubyear{2017}

\begin{document}

\label{firstpage}
\pagerange{\pageref{firstpage}--\pageref{lastpage}}
\maketitle

\begin{abstract}
We calculated the polarization degree of hydrogen Balmer broad emission lines from a number of active galactic nuclei (AGNs) with determined virial factors. The objects were selected from the sample presented by \citet{decarli08}. In our calculations, we used the model of the flattened disc-like structure of the broad-line emission region (BLR). In this model, the expression for the virial factor makes it possible to determine the inclination angle for the flattened BLR, which in turn yields the polarization degree of the broad emission lines. As a result, we obtained the direct relation between the polarization degree and the virial factor. We also compared the determined values of the polarization degree with those obtained in polarimetric observations.
\end{abstract}


\begin{table*}
\centering
\caption{The dependence of the degree of polarization on the virial factor $f = V_{BLR} / FWHM$ with the absorption ratio $q=0.0,0.01,0.05,0.1,0.2,0.3$ for objects from \citet{decarli08}.}
\begin{tabular}{llllllll}
\hline
\textbf{Object}&
\textbf{$f \backslash q$}&
\textbf{0.0}&
\textbf{0.01}&
\textbf{0.05}&
\textbf{0.1}&
\textbf{0.2}&
\textbf{0.3} \\
\hline
0054+144&
0.70&
1.11&
1.58&
3.40&
5.58&
9.59&
13.20 \\

0100+0205&
0.80&
0.76&
1.12&
2.51&
4.15&
7.13&
9.78 \\

0110+297&
1.10&
0.35&
0.54&
1.26&
2.11&
3.62&
4.93 \\

0133+207&
0.60&
1.89&
2.53&
5.04&
8.08&
13.79&
19.08 \\

3C48&
3.00&
0.04&
0.07&
0.17&
0.29&
0.50&
0.67 \\

0204+292&
1.10&
0.35&
0.54&
1.26&
2.11&
3.62&
4.93 \\

0244+194&
1.10&
0.35&
0.54&
1.26&
2.11&
3.62&
4.93 \\

0624+6907&
1.80&
0.12&
0.19&
0.46&
0.78&
1.33&
1.80 \\

07546+3928&
4.60&
0.01&
0.02&
0.06&
0.10&
0.17&
0.22 \\

US1867&
2.60&
0.06&
0.10&
0.23&
0.39&
0.67&
0.90 \\

0944.1+1333&
3.40&
0.03&
0.05&
0.12&
0.19&
0.33&
0.45 \\

0953+415&
0.70&
1.11&
1.58&
3.40&
5.58&
9.59&
13.20 \\

1001+291&
4.40&
0.01&
0.02&
0.06&
0.10&
0.17&
0.22 \\

1004+130&
0.90&
0.57&
0.85&
1.94&
3.23&
5.55&
7.59 \\

1058+110&
0.80&
0.76&
1.12&
2.51&
4.15&
7.13&
9.78 \\

1100+772&
0.60&
1.89&
2.53&
5.04&
8.08&
13.79&
19.08 \\

1150+497&
3.10&
0.04&
0.07&
0.17&
0.29&
0.50&
0.67 \\

1202+281&
1.20&
0.28&
0.44&
1.04&
1.73&
2.97&
4.04 \\

1216+069&
0.50&
11.71&
12.59&
16.07&
20.35&
28.64&
36.64 \\

Mrk 0205&
2.20&
0.07&
0.12&
0.29&
0.49&
0.83&
1.12 \\

1222+125&
1.00&
0.44&
0.67&
1.55&
2.58&
4.43&
6.04 \\

1230+097&
1.70&
0.14&
0.22&
0.52&
0.87&
1.49&
2.02 \\

1307+085&
0.70&
1.11&
1.58&
3.40&
5.58&
9.59&
13.20 \\

1309+355&
2.20&
0.07&
0.12&
0.29&
0.49&
0.83&
1.12 \\

1402+436&
1.50&
0.17&
0.26&
0.64&
1.07&
1.82&
2.47 \\

1425+267&
0.70&
1.11&
1.58&
3.40&
5.58&
9.59&
13.20 \\

1444+407&
1.80&
0.12&
0.19&
0.46&
0.78&
1.33&
1.80 \\

1512+37&
0.70&
1.11&
1.58&
3.40&
5.58&
9.59&
13.20 \\

3C323.1&
1.10&
0.35&
0.54&
1.26&
2.11&
3.62&
4.93 \\

1549+203&
2.40&
0.06&
0.10&
0.23&
0.39&
0.67&
0.90 \\

1635+119&
0.60&
1.89&
2.53&
5.04&
8.08&
13.79&
19.08 \\

3C351&
0.50&
11.71&
12.59&
16.07&
20.35&
28.64&
36.64 \\

1821+643&
1.30&
0.23&
0.36&
0.87&
1.45&
2.48&
3.36 \\

2141+175&
1.30&
0.23&
0.36&
0.87&
1.45&
2.48&
3.36 \\

2201+315&
3.20&
0.03&
0.05&
0.12&
0.19&
0.33&
0.45 \\

2247+140&
2.70&
0.04&
0.07&
0.17&
0.29&
0.50&
0.67 \\
\hline
\end{tabular}
\label{tab1}
\end{table*}

\begin{table*}
\begin{center}
\caption{Comparison of determined and observed polarization degrees for AGNs from \citet{decarli08} sample. Columns: 1. Object; 2. Redshift; 3. The virial product; 4. The virial factor determined by Eq.(\ref{eq01}); 5. The BLR inclination angle obtained from Eq.(\ref{eq03}); 6. $\delta i = i_{BLR} - i_P$, where $i_P$ is the inclination angle determined from polarimetric data; 7. The ratio $H / R$ for the BLR disk-like structure according to Eq.(\ref{eq02}); 8. The polarization degree of BLR emission; 9. Wide band polarization from the observed data.  Observational data are taken from (1) \citet{wills92}, (2) \citet{afanasiev11}, (3) \citet{berriman90}, (4) \citet{takalo92}, (5) \citet{hines01}.}
\begin{tabular}{llcllllll}
\hline
{\bf Object} & $z$ & $\log{VP(H_\beta)}$& $f$ & $i_{BLR}^\circ$ & $\Delta i$& $\left({\frac{H}{R}}\right)_{BLR} $& $P_l [\%]$ & $P_l [\%]$ \\
             &     & $[M_\odot]$        &     &                 &           &                                    &  BLR       & Obs \\
\hline
0054+144&
0.171&
8.9&
0.49&
45.6&
&
&
1.11&
 \\

0100+0205&
0.393&
8.6&
0.64&
38.7&
&
&
0.76&
 \\

01110+297&
0.363&
8.6&
1.21&
27.1&
36.9&
&
0.35&
2.6$^{(1)}$ \\

0133+207&
0.424&
9.0&
0.36&
53&
&
0.241&
1.62&
1.62$^{(1)}$ \\

3C48&
0.368&
8.7&
9.0&
9.6&
30&
&
0.04&
0.99$^{(1)}$ \\

0204+292&
0.110&
8.6&
1.21&
27.1&
&
&
0.35&
 \\

0244+194&
0.174&
8.4&
1.21&
27.1&
&
&
0.35&
 \\

0624+6907&
0.370&
9.1&
3.24&
16.2&
&
&
0.12&
 \\

07546+3928&
0.096&
8.0&
21.16&
6.3&
&
&
0.01&
 \\

US1867&
0.515&
8.2&
6.76&
11&
&
&
0.06&
 \\

09441+1333&
0.134&
7.9&
11.56&
8.5&
&
&
0.03&
 \\

0953+415&
0.235&
8.8&
0.49&
29&
&
0.53&
0.4&
0.39$\pm $0.12$^{(2)}$ \\

1001+291&
0.330&
7.8&
19.36&
6.6&
32.4&
&
0.22&
$0.77\pm 0.22^{(3)}$ \\

1004+130&
0.241&
9.0&
0.81&
33.8&
&
&
0.56&
0.51$\pm $0.14$^{(3)}$ \\

1058+110&
0.423&
8.7&
0.64&
38.7&
&
&
1.12&
1.01$^{(1)}$ \\

1100+772&
0.311&
9.5&
0.36&
38&
&
0.57&
0.71&
0.71$\pm $0.22$^{(3)}$ \\

1150+497&
0.334&
8.1&
9.61&
9.3&
31&
&
0.03&
0.84$^{(1)}$ \\

1202+281&
0.165&
8.4&
1.44&
24.6&
&
&
0.28&
0.34$\pm $0.15$^{(3)}$ \\

1216+069&
0.331&
9.1&
0.25&
39.6&
&
0.59&
0.80&
0.80$\pm $0.19$^{(3)}$ \\

Mrk 0205&
0.071&
8.0&
4.84&
13.1&
&
&
0.07&
$0.13\pm 0.15 ^{(4)}$ \\

1222+125&
0.412&
8.8&
1.0&
30&
&
&
0.41&
 \\

1230+097&
0.416&
8.8&
2.89&
17.1&
&
&
0.14&
 \\

1307+085&
0.155&
8.5&
0.49&
31.5&
&
0.5&
0.48&
$0.48\pm 0.16^{(3)}$ \\

1309+355&
0.184&
8.1&
4.84&
26&
&
&
0.3 (q=0.2)&
0.31$\pm $0.14$^{(3)}$ \\

1402+436&
0.323&
8.5&
2.25&
19.5&
&
&
1.86&
1.86$\pm $0.30$^{(5)}$ \\

1425+267&
0.366&
9.2&
0.49&
45.6&
&
&
1.11&
1.63$\pm $0.47$^{(2)}$ \\

1444+407&
0.267&
8.2&
3.24&
16.2&
11.6&
&
0.39 (q=0.05)&
0.37$\pm $0.15$^{(3)}$ \\

1512+637&
0.371&
9.2&
0.49&
45.6&
&
&
1.11&
$1.10\pm 0.22^{(3)}$ \\

1545+210&
0.266&
8.8&
1.21&
27.1&
26.4&
&
1.26 (q=0.05)&
1.51$\pm $0.30$^{(3)}$ \\

1549+203&
0.253&
7.5&
5.76&
12&
&
&
0.06&
 \\

1635+119&
0.148&
6.9&
0.36&
35.2&
&
0.602&
0.6&
0.61$^{(1)}$ \\

1704+608&
0.372&
9.6&
0.25&
26.5&
&
0.9&
0.33&
0.31$\pm $0.17$^{(3)}$ \\

1821+623&
0.297&
9.4&
1.69&
22.7&
&
&
0.23&
 \\

2141+175&
0.211&
8.7&
1.69&
22.7&
&
&
0.23&
0.22$^{(1)}$ \\

2201+315&
0.295&
8.4&
10.24&
9.0&
18.5&
&
0.36 (q=0.2)&
0.72$\pm $0.57$^{(1)}$ \\

2247+140&
0.235&
8.0&
7.29&
10.7&
31.3&
&
0.7 (q=0.03)&
0.92$^{(1)}$ \\
\hline
\end{tabular}
\label{tab2}
\end{center}
\end{table*}

\section{Introduction}

Accretion discs exist around supermassive black holes (SMBHs) in active galactic nuclei (AGNs). The broad optical/UV emission lines in the spectra of Seyfert 1 galaxies and quasars are produced at the distances of about 1 to 100 light days from the central black hole. The regions that emit these lines are spatially unresolved, even for the nearest AGNs. The basic information about the sizes of the broad-line emitting regions (BLR) is generally obtained from the delayed variability of the intensities of integrated emission lines with respect to the continuum radiation. This method is usually called ''reverberation mapping'' \citep{cherepashchuk73}. BLRs are stratified: lines from highly ionized species originate closer to the central ionizing source than do lines from lower ions \citep{gaskell86,krolik91,peterson00,kollatschny03,kollatschny13}.

The main problem is the origin of the structure and kinematics of BLRs. \citet{gaskell13} suggested that a BLR is a part of the outer accretion disc and that similar MHD processes occur. A disc-like geometry for a BLR was proposed by several authors \citep{labita06,decarli08}. Some authors suggest that the BLR cannot be completely flat \citep{collin06}, which implies that a disc may have a finite half-thickness $H$, or a profile with $H$ increasing with the disc radius. Some models suggest the existence of warped discs \citep{tremaine14}.

Properties of the broad emission lines can be used to estimate the mass of the central SMBH. The traditional method is based on the virial theorem: the mass of SMBH is estimated from the following relation \citep{vestergaard06}:

\begin{equation}
 M_{BH} = f \frac{R_{BLR} V_{BLR}^2}{G},
 \label{eq01}
\end{equation}

\noindent where $M_{BH}$ is the mass of the black hole, $R_{BLR}$ is the radius of BLR, $V_{BLR}$ is the velocity of accreting gas in BLR, which is usually measured as the full width of the emission line in the BLR emission spectrum, and $f$ is the virial factor that defines the geometry, velocity field, and orientation of the BLR. Thereby, the virial factor $f$ depends on the inclination angle $i$ of the BLR. According to \citet{collin06} and \citet{decarli11}, the expression for the virial factor can be presented in the following form:

\begin{figure}
	\includegraphics[width=\columnwidth]{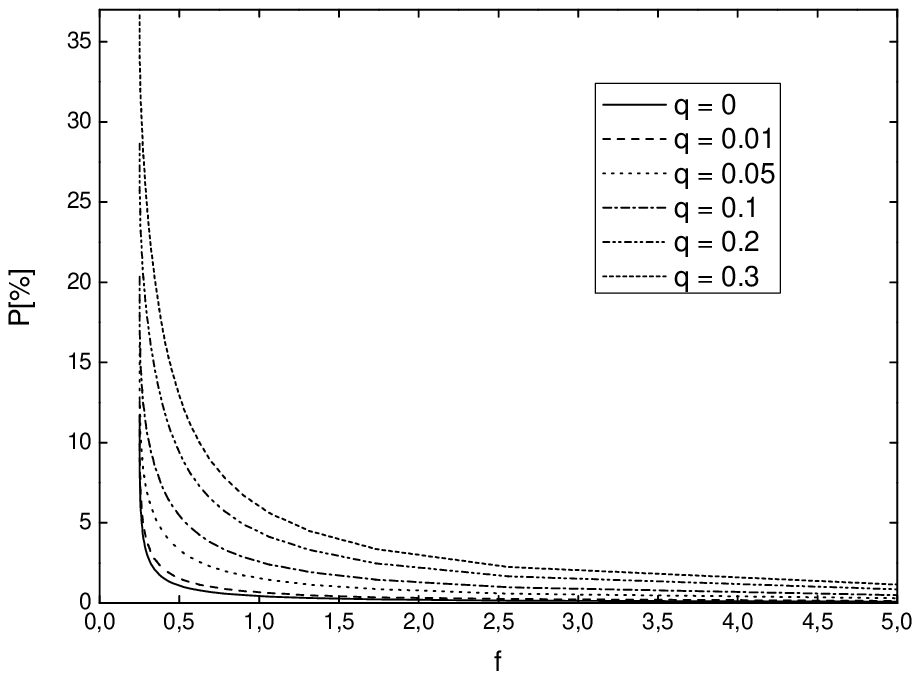}
  \caption{Dependence of the degree of polarization on the virial factor $f$ and absorption ratio $q$.}
  \label{fig01}
\end{figure}

\begin{equation}
 f = 0.25 \left[\left(\frac{H}{R}\right)^2 + sin^2{i}\right]^{-1},\,\, V_{BLR} = FWHM,
 \label{eq02}
\end{equation}

\noindent where $H/R$ is the aspect ratio of the disc. For a geometrically thin disc, the equation (\ref{eq02}) reduces to \citep{mclure01}:

\begin{equation}
 f = \frac{1}{4 sin^2{i}}.
 \label{eq03}
\end{equation}

\noindent Within these limits, the virial factor $f$ ranges from 0.5 (edge-on discs) to $\infty$ (face-on discs). For small inclination angles, when $i \rightarrow 0$, $f \rightarrow 0.25 (H/R)^{-2}$ and $f = 1$ for $H = 0.5 R$.

There are a number of papers \citep{decarli08,decarli11,labita08}, in which the virial factors were determined for SMBHs in the sufficiently large number of AGNS.

Our sample of AGN includes the targets for which the virial factor is determined by authors of publications (Table \ref{tab1}) and the targets for which the polarization of broad line emission and continuum was measured (Table \ref{tab2}).

Here, we determine the polarization degree of the radiation of BLR of AGNs from the lists of \citet{decarli08,decarli11} and \citet{labita08}. To estimate the degree of polarization, we used the equation (\ref{eq03}) which allows us to obtain the value of the inclination angle $i$ directly from the polarization degree of the radiation of the disc-like structure, on the basis of the widely accepted theory developed by \citet{chanrdasekhar50} and \citet{sobolev63}. For a disc-like structure, the scattered radiation displays the maximum linear polarization $P_l = 11.7\%$ when the line of sight is perpendicular to the normal to the semi-infinite atmosphere (Milne problem).

\citet{chanrdasekhar50} and \citet{sobolev63} presented the solution for the so-called Milne problem. Their solution corresponds to multiple scattering of light in optically thick flattened atmospheres and can be applied to BLR. The idea of inferring the inclination of the accretion disc around a black hole from observations of its polarized continuum was suggested by \citet{li09}.

The values of polarization $P_l(\mu)$, where $\mu = \cos{i}$, are presented in the detailed table from our earlier paper \citet{gnedin15}. The table also presents the dependence of the polarization degree on the absorption degree $q = \sigma_a / (\sigma_s + \sigma_a)$, where $\sigma_a$ is the absorption cross-section, and $\sigma_s$ is the cross-section of scattering.

Using the data from \citet{gnedin15} we presented at Fig.1 the dependence of the polarization degree on the virial factor according to Eq.(\ref{eq03}), taking into account the effect of absorption in BLR. It is very important that the effect of absorption provides the increase in the polarization degree.

Polarization of radiation emitted from the plane-parallel medium is determined by the anisotropy of the radiation. If the radiation is isotropic it is unpolarized. In the case of pure scattering ($q = 0.0$) the multiple scattered radiation becomes more and more isotropic with the distance from the source of radiation. Absorption increases the anisotropy of outgoing radiation and hence increases the polarization degree (see details in \citet{dolginov95}). Fig.\ref{fig01} demonstrates the increase in polarization degree with the increase in the degree of absorption $q$.

We calculated the polarization degree for a number of AGNs for which the virial factor had been obtained.

The basic problem is whether it is possible to apply the Chanrasekhar-Sobolev theory of multiple scattering to BLR radiation. The key issue is the optical thickness and geometry of BLR. According to \citet{kollatschny13}, the BLR is a geometrically thick disk with a thickness-to-radius ratio $H / R < 1.0$. \citet{kollatschny13} estimated the ratio $H / R$ for some AGNs. For example, for NGC~5548 this estimate is $H /R \approx 0.1$. It allows to estimate the optical thickness of BLR, using the value of the gas density near the inner radius of BLR as $N_H = 10^{11.5} cm^{-3}$ \citep{poutanen10}. \citet{kollatschny13} considered in detail the vertical BLR structure in AGN and estimated the BLR radius for NGC~5548 as $R_{BLR} = 20$ light days. With this value, we estimated the optical thickness of BLR for the Thomson scattering as follows: $\tau = N_H \sigma_{Th} H = 10^3$. It means that the BLR is really a disk-like, optically thick structure and one can use Chanrdasekhar-Sobolev theory to estimate the polarization degree of BLR emission.

\section{The polarization of radiation from geometrically thin BLR for quasars with determined virial factors}

\citet{decarli08} presented the values of the virial product and the virial factor for a number of quasars (see Table 5 from their paper). The virial products are estimated from the luminosity of the host galaxy, namely, on the basis of $M_{BH} - L_{bulge}$ relation. The mean value of the virial factor is $f^{0.5} = 1.6 \pm 1.1$. The real values of $f$ for the specific quasars are presented in the column 4 of Table 5 of \citet{decarli08}. Then, using Eq.(\ref{eq03}) we obtained the values of $\sin{i}$ and the inclination angle $i$ itself. Using the data from Table 1 from our paper \citet{gnedin15}, we estimated the polarization degree dependent on $\mu = \cos{i}$ and the absorption ratio $q$. Eq.(\ref{eq03}) makes it possible to determine the dependence of the polarization degree of broad emission lines on the virial factor. This dependence is presented in Fig.\ref{fig01}.

Note that we use the virial factor $f$ from Table 5 of \citet{decarli08}, where the virial product ($M_{BH}$) rather than the BH mass is presented, and the latter is derived from the absolute magnitude of the bulge component. We should emphasize that in our Table \ref{tab2} we use only the virial factor $f$, but not the real value of $M_{BH}$. Thus, we show that polarimetric observations make it possible to estimate independently the virial factor $f$, which is commonly used to estimate $M_{BH}$.

The final results of our calculations are presented in the Table \ref{tab1}. The quasars with determined virial factor $f$ \citep{decarli08} are listed in the first column of this Table. It should be emphasized that \citet{decarli08} determined the virial factors as $V_{BLR} = f FWHM$. Thereby, the virial factors from Eq.(\ref{eq03}) and from \citet{decarli08} are connected via the relation $f(Eq.3) = f^2(Decarli)$.

\section{Comparison of the determined polarization degree with the observed data}

It is important to compare the determined polarization degree with the published data of polarimetric observations. Unfortunately, the polarimetric data for the objects with determined virial factor presented by \citet{decarli08} (Table 1 in our paper) were obtained in broad bands with the concentration near V band \citep{berriman90,wills92,afanasiev11}. The comparison is presented in Table \ref{tab2}, which demonstrates that our estimates for the polarization degree coincide with the observed data for 20 AGNs. This fact may indicate the equality between inclination angles of BLR and that of the standard accretion disc, i.e. $i_{BLR} = i_a$, where $i_a$ is the inclination angle of the accretion disc. For a number of objects, such as 0133+207, 0953+415, 1100+772, 1216+069, 1307+085, 1635+119 and 1704+608 it is found that the BLRs of these objects are geometrically thick discs, in accordance with Eq.(\ref{eq02}).

The information about the geometrically thick BLR can be obtained with Eqs.(\ref{eq01})-(\ref{eq03}). The real value of the virial factor can be obtained from Eq.(\ref{eq01}) with the BH mass presented in the column 3 of our Table \ref{tab2}. The data on $R_{BLR}$ and $V_{BLR}$ can be obtained from \citet{decarli08} and \citet{decarli11}. To estimate the geometrical thickness of BLR, we used Eq.(\ref{eq02}). In this case the virial factor $f$ depends on $\sin{i}$, which in turn can be obtained from the measured polarization degrees using Chandrasekhar-Sobolev theory of generation of polarization due to multiple scattering in an optically thick disk of the same type as the objects considered.

\section{Conclusion}

We calculated the expected polarization of hydrogen broad-line emission for the sample of AGNs presented by \citet{decarli08}, under the assumption of the disc-like structure of the BLR; both geometrically thin and geometrically thick discs were considered. If the disc of BLR is geometrically thin, the virial factor takes a form $f = (2 \sin{i})^{-2}$. For the geometrically thick disc-like structure, the virial factor is derived via Eq.(\ref{eq02}). For a number of objects, BLRs are geometrically thick disc-like structures. This fact confirms the conclusion that $H_{\alpha,\beta}$ lines are emitted in a more distant region than are UV lines (CIV, MgII) and that the disc is thinner in the inner region, where CIV and MgII lines are emitted \citep{decarli08}. The virial factor $f$ clearly depends on the width of broad lines, which implies that for a geometrically thin disc-like structure of BLR, the polarization degree of the line emission and the width of this line strongly correlate. For a number of objects, our estimated values of polarization degree are consistent with those observed (Table \ref{tab2}). For a number of other objects, systematic polarimetric observations are still needed.

\section*{Acknowledgements}

This research was supported by the Basic Research Program P-7 of Praesidium of Russian Academy od Sciences and the program of the Department of Physical Sciences of Russian Academy of Sciences No.2. Yu.N. Gnedin and S.D. Buliga were supported by the presidential program ''Leading Scientific School-7241.2016.2''.

\bibliographystyle{mnras}
\bibliography{mybibfile}

\begin{thebibliography}{}
\makeatletter
\relax
\def\mn@urlcharsother{\let\do\@makeother \do\$\do\&\do\#\do\^\do\_\do\%\do\~}
\def\mn@doi{\begingroup\mn@urlcharsother \@ifnextchar [ {\mn@doi@}
  {\mn@doi@[]}}
\def\mn@doi@[#1]#2{\def\@tempa{#1}\ifx\@tempa\@empty \href
  {http://dx.doi.org/#2} {doi:#2}\else \href {http://dx.doi.org/#2} {#1}\fi
  \endgroup}
\def\mn@eprint#1#2{\mn@eprint@#1:#2::\@nil}
\def\mn@eprint@arXiv#1{\href {http://arxiv.org/abs/#1} {{\tt arXiv:#1}}}
\def\mn@eprint@dblp#1{\href {http://dblp.uni-trier.de/rec/bibtex/#1.xml}
  {dblp:#1}}
\def\mn@eprint@#1:#2:#3:#4\@nil{\def\@tempa {#1}\def\@tempb {#2}\def\@tempc
  {#3}\ifx \@tempc \@empty \let \@tempc \@tempb \let \@tempb \@tempa \fi \ifx
  \@tempb \@empty \def\@tempb {arXiv}\fi \@ifundefined
  {mn@eprint@\@tempb}{\@tempb:\@tempc}{\expandafter \expandafter \csname
  mn@eprint@\@tempb\endcsname \expandafter{\@tempc}}}

\bibitem[\protect\citeauthoryear{{Afanasiev}, {Borisov}, {Gnedin},
  {Natsvlishvili}, {Piotrovich}  \& {Buliga}}{{Afanasiev}
  et~al.}{2011}]{afanasiev11}
{Afanasiev} V.~L.,  {Borisov} N.~V.,  {Gnedin} Y.~N.,  {Natsvlishvili} T.~M.,
  {Piotrovich} M.~Y.,   {Buliga} S.~D.,  2011, \mn@doi [Astronomy Letters]
  {10.1134/S106377371105001X}, \href
  {http://adsabs.harvard.edu/abs/2011AstL...37..302A} {37, 302}

\bibitem[\protect\citeauthoryear{{Berriman}, {Schmidt}, {West}  \&
  {Stockman}}{{Berriman} et~al.}{1990}]{berriman90}
{Berriman} G.,  {Schmidt} G.~D.,  {West} S.~C.,   {Stockman} H.~S.,  1990,
  \mn@doi [\apjs] {10.1086/191523}, \href
  {http://adsabs.harvard.edu/abs/1990ApJS...74..869B} {74, 869}

\bibitem[\protect\citeauthoryear{{Chandrasekhar}}{{Chandrasekhar}}{1950}]{chanrdasekhar50}
{Chandrasekhar} S.,  1950, {Radiative transfer.}

\bibitem[\protect\citeauthoryear{{Cherepashchuk} \& {Lyutyi}}{{Cherepashchuk}
  \& {Lyutyi}}{1973}]{cherepashchuk73}
{Cherepashchuk} A.~M.,  {Lyutyi} V.~M.,  1973, \aplett, \href
  {http://adsabs.harvard.edu/abs/1973ApL....13..165C} {13, 165}

\bibitem[\protect\citeauthoryear{{Collin}, {Kawaguchi}, {Peterson}  \&
  {Vestergaard}}{{Collin} et~al.}{2006}]{collin06}
{Collin} S.,  {Kawaguchi} T.,  {Peterson} B.~M.,   {Vestergaard} M.,  2006,
  \mn@doi [\aap] {10.1051/0004-6361:20064878}, \href
  {http://adsabs.harvard.edu/abs/2006A%26A...456...75C} {456, 75}

\bibitem[\protect\citeauthoryear{{Decarli}, {Labita}, {Treves}  \&
  {Falomo}}{{Decarli} et~al.}{2008}]{decarli08}
{Decarli} R.,  {Labita} M.,  {Treves} A.,   {Falomo} R.,  2008, \mn@doi
  [\mnras] {10.1111/j.1365-2966.2008.13320.x}, \href
  {http://adsabs.harvard.edu/abs/2008MNRAS.387.1237D} {387, 1237}

\bibitem[\protect\citeauthoryear{{Decarli}, {Dotti}  \& {Treves}}{{Decarli}
  et~al.}{2011}]{decarli11}
{Decarli} R.,  {Dotti} M.,   {Treves} A.,  2011, \mn@doi [\mnras]
  {10.1111/j.1365-2966.2010.18102.x}, \href
  {http://adsabs.harvard.edu/abs/2011MNRAS.413...39D} {413, 39}

\bibitem[\protect\citeauthoryear{{Dolginov}, {Gnedin}  \&
  {Silant'ev}}{{Dolginov} et~al.}{1995}]{dolginov95}
{Dolginov} A.~Z.,  {Gnedin} Y.~N.,   {Silant'ev} N.~A.,  1995, {Propagation and
  polarisation of radiation in cosmic media}

\bibitem[\protect\citeauthoryear{{Gaskell} \& {Goosmann}}{{Gaskell} \&
  {Goosmann}}{2013}]{gaskell13}
{Gaskell} C.~M.,  {Goosmann} R.~W.,  2013, \mn@doi [\apj]
  {10.1088/0004-637X/769/1/30}, \href
  {http://adsabs.harvard.edu/abs/2013ApJ...769...30G} {769, 30}

\bibitem[\protect\citeauthoryear{{Gaskell} \& {Sparke}}{{Gaskell} \&
  {Sparke}}{1986}]{gaskell86}
{Gaskell} C.~M.,  {Sparke} L.~S.,  1986, \mn@doi [\apj] {10.1086/164238}, \href
  {http://adsabs.harvard.edu/abs/1986ApJ...305..175G} {305, 175}

\bibitem[\protect\citeauthoryear{{Gnedin}, {Piotrovich}, {Silant'ev},
  {Natsvlishvili}  \& {Buliga}}{{Gnedin} et~al.}{2015}]{gnedin15}
{Gnedin} Y.~N.,  {Piotrovich} M.~Y.,  {Silant'ev} N.~A.,  {Natsvlishvili}
  T.~M.,   {Buliga} S.~D.,  2015, \mn@doi [Astrophysics]
  {10.1007/s10511-015-9398-1}, \href
  {http://adsabs.harvard.edu/abs/2015Ap.....58..443G} {58, 443}

\bibitem[\protect\citeauthoryear{{Hines}, {Schmidt}, {Gordon}, {Smith},
  {Wills}, {Allen}  \& {Sitko}}{{Hines} et~al.}{2001}]{hines01}
{Hines} D.~C.,  {Schmidt} G.~D.,  {Gordon} K.~D.,  {Smith} P.~S.,  {Wills}
  B.~J.,  {Allen} R.~G.,   {Sitko} M.~L.,  2001, \mn@doi [\apj]
  {10.1086/323954}, \href {http://adsabs.harvard.edu/abs/2001ApJ...563..512H}
  {563, 512}

\bibitem[\protect\citeauthoryear{{Kollatschny}}{{Kollatschny}}{2003}]{kollatschny03}
{Kollatschny} W.,  2003, \mn@doi [\aap] {10.1051/0004-6361:20030928}, \href
  {http://adsabs.harvard.edu/abs/2003A%26A...407..461K} {407, 461}

\bibitem[\protect\citeauthoryear{{Kollatschny} \& {Zetzl}}{{Kollatschny} \&
  {Zetzl}}{2013}]{kollatschny13}
{Kollatschny} W.,  {Zetzl} M.,  2013, \mn@doi [\aap]
  {10.1051/0004-6361/201321685}, \href
  {http://adsabs.harvard.edu/abs/2013A%26A...558A..26K} {558, A26}

\bibitem[\protect\citeauthoryear{{Krolik}, {Horne}, {Kallman}, {Malkan},
  {Edelson}  \& {Kriss}}{{Krolik} et~al.}{1991}]{krolik91}
{Krolik} J.~H.,  {Horne} K.,  {Kallman} T.~R.,  {Malkan} M.~A.,  {Edelson}
  R.~A.,   {Kriss} G.~A.,  1991, \mn@doi [\apj] {10.1086/169918}, \href
  {http://adsabs.harvard.edu/abs/1991ApJ...371..541K} {371, 541}

\bibitem[\protect\citeauthoryear{{Labita}, {Treves}, {Falomo}  \&
  {Uslenghi}}{{Labita} et~al.}{2006}]{labita06}
{Labita} M.,  {Treves} A.,  {Falomo} R.,   {Uslenghi} M.,  2006, \mn@doi
  [\mnras] {10.1111/j.1365-2966.2006.10878.x}, \href
  {http://adsabs.harvard.edu/abs/2006MNRAS.373..551L} {373, 551}

\bibitem[\protect\citeauthoryear{{Labita}, {Treves}  \& {Falomo}}{{Labita}
  et~al.}{2008}]{labita08}
{Labita} M.,  {Treves} A.,   {Falomo} R.,  2008, \mn@doi [\mnras]
  {10.1111/j.1365-2966.2007.12656.x}, \href
  {http://adsabs.harvard.edu/abs/2008MNRAS.383.1513L} {383, 1513}

\bibitem[\protect\citeauthoryear{{Li}, {Narayan}  \& {McClintock}}{{Li}
  et~al.}{2009}]{li09}
{Li} L.-X.,  {Narayan} R.,   {McClintock} J.~E.,  2009, \mn@doi [\apj]
  {10.1088/0004-637X/691/1/847}, \href
  {http://adsabs.harvard.edu/abs/2009ApJ...691..847L} {691, 847}

\bibitem[\protect\citeauthoryear{{McLure} \& {Dunlop}}{{McLure} \&
  {Dunlop}}{2001}]{mclure01}
{McLure} R.~J.,  {Dunlop} J.~S.,  2001, \mn@doi [\mnras]
  {10.1046/j.1365-8711.2001.04709.x}, \href
  {http://adsabs.harvard.edu/abs/2001MNRAS.327..199M} {327, 199}

\bibitem[\protect\citeauthoryear{{Peterson} \& {Wandel}}{{Peterson} \&
  {Wandel}}{2000}]{peterson00}
{Peterson} B.~M.,  {Wandel} A.,  2000, \mn@doi [\apjl] {10.1086/312862}, \href
  {http://adsabs.harvard.edu/abs/2000ApJ...540L..13P} {540, L13}

\bibitem[\protect\citeauthoryear{{Poutanen} \& {Stern}}{{Poutanen} \&
  {Stern}}{2010}]{poutanen10}
{Poutanen} J.,  {Stern} B.,  2010, \mn@doi [\apjl]
  {10.1088/2041-8205/717/2/L118}, \href
  {http://adsabs.harvard.edu/abs/2010ApJ...717L.118P} {717, L118}

\bibitem[\protect\citeauthoryear{{Sobolev}}{{Sobolev}}{1963}]{sobolev63}
{Sobolev} V.~V.,  1963, {A treatise on radiative transfer.}

\bibitem[\protect\citeauthoryear{{Takalo}, {Kidger}, {de Diego}  \&
  {Sillanpaa}}{{Takalo} et~al.}{1992}]{takalo92}
{Takalo} L.~O.,  {Kidger} M.~R.,  {de Diego} J.~A.,   {Sillanpaa} A.,  1992,
  \aap, \href {http://adsabs.harvard.edu/abs/1992A%26A...261..415T} {261, 415}

\bibitem[\protect\citeauthoryear{{Tremaine} \& {Davis}}{{Tremaine} \&
  {Davis}}{2014}]{tremaine14}
{Tremaine} S.,  {Davis} S.~W.,  2014, \mn@doi [\mnras] {10.1093/mnras/stu663},
  \href {http://adsabs.harvard.edu/abs/2014MNRAS.441.1408T} {441, 1408}

\bibitem[\protect\citeauthoryear{{Vestergaard} \& {Peterson}}{{Vestergaard} \&
  {Peterson}}{2006}]{vestergaard06}
{Vestergaard} M.,  {Peterson} B.~M.,  2006, \mn@doi [\apj] {10.1086/500572},
  \href {http://adsabs.harvard.edu/abs/2006ApJ...641..689V} {641, 689}

\bibitem[\protect\citeauthoryear{{Wills}, {Wills}, {Breger}, {Antonucci}  \&
  {Barvainis}}{{Wills} et~al.}{1992}]{wills92}
{Wills} B.~J.,  {Wills} D.,  {Breger} M.,  {Antonucci} R.~R.~J.,   {Barvainis}
  R.,  1992, \mn@doi [\apj] {10.1086/171869}, \href
  {http://adsabs.harvard.edu/abs/1992ApJ...398..454W} {398, 454}

\makeatother
\end{thebibliography}

\bsp
\label{lastpage}
\end{document}